# Probing the Role of Magnetic-Field Variations in NOAA AR 8038 in Producing Solar Flare and CME on 12 May 1997

Rajmal Jain • Arun K. Awasthi • Babita Chandel • Lokesh Bharti • Y. Hanaoka • A. L. Kiplinger

**Abstract** We carried out a multi-wavelength study of a Coronal Mass Ejection (CME) and an associated flare occurring on 12 May 1997. We present a detailed investigation of magnetic-field variations in the NOAA Active Region 8038 which was observed on the Sun during 7--16 May 1997. This region was quiet and decaying and produced only very small flare activity during its disk passage. However, on 12 May 1997 it produced a CME and associated medium-size 1B/C1.3 flare. Detailed analyses of Hα filtergrams and MDI/SOHO magnetograms revealed continual but discrete surge activity, and emergence and cancellation of flux in this active region. The movie of these magnetograms revealed two important results that the major opposite polarities of pre-existing region as well as in the emerging flux region were approaching towards each other and moving magnetic features (MMF) were ejecting out from the major north polarity at a quasi-periodicity of about ten hrs during 10--13 May 1997. These activities were probably caused by the magnetic reconnection in the lower atmosphere driven by photospheric convergence motions, which were evident in magnetograms. The quantitative measurements of magnetic field variations such as magnetic flux, gradient, and sunspot rotation revealed that in this active region, free energy was slowly being stored in the corona. The slow low-layer magnetic reconnection may be responsible for the storage of magnetic free energy in the corona and the formation of a sigmoidal core field or a flux rope leading to the eventual

Rajmal Jain (✉) • Arun K. Awasthi
Physical Research Laboratory, Dept. of Space, Navrangpura, Ahmedabad - 380009, India
e-mail: rajmal@prl.res.in

*Present address:*
Babita Chandel
Dept. Of Physics, Manav Bharti University, Solan, H.P., India

Lokesh Bharti
Max-Planck-Institute for Solar System Research, Max-Planck-Str. 2, 37191 Katlenburg-Lindau, Germany

Y. Hanaoka
National Astronomical Observatory of Japan, 2-21-1 Osawa, Mitaka, Tokyo 181-8588, Japan

A. L. Kiplinger
Center for Integrated Plasma Studies, University of Colorado, Boulder, CO 80309, USA

eruption. The occurrence of EUV brightenings in the sigmoidal core field prior to the rise of a flux rope suggests that the eruption was triggered by the inner tether-cutting reconnection, but not the external breakout reconnection. An impulsive acceleration revealed from fast separation of the Hα ribbons of the first 150 seconds suggests the CME accelerated in the inner corona, which is also in consistent with the temporal profile of the reconnection electric field. Based on observations and analysis we propose a qualitative model, and we conclude that the mass ejections, filament eruption, CME, and subsequent flare were connected with one another and should be regarded within the framework of a solar eruption.

**Keywords** Sun: magnetograms • Sun: flares • Sun: CMEs

## 1. Introduction

It has been widely accepted that the energy released at the time of the flare and Coronal Mass Ejection (CME) in an active region is derived from the gradually stored energy of the surrounding magnetic fields which are in a non-potential state and twisted or sheared magnetic loops (Jain, 1983; Martens and Kuin, 1989; Aurass *et al.*, 1999; Schmieder, 2006). However, the observations do not show any drastic change in the magnetic field at the time of the flare in an active region. Rather they reveal that stresses in the coronal magnetic fields may build up in response to the changes taking place at the photospheric level, such as sunspot motion and emerging fluxes. During the evolution of an active region and passage over the disk many phenomena are observed *viz.* the motion of an active region around its own axis, moving magnetic features (MMFs), cancellation of active region fluxes, new flux emergence, plage brightening, surge activity, and filament eruptions. Magnetic-field variations have now been well quantified and studied to probe their relationship with energetic phenomena *viz.* flares, eruptive prominences, and CMEs.

One way to quantify shear in an active region is to measure rotation of the leading and following polarity sunspots around their center of mass known as "Rotation angle" (Zhang *et al.*, 2008). The continuous ongoing processes of emergence and cancellation of flux changes the total flux of the active region, which may be quantified by the product of magnetic field strength [H] and area [A] of the region $\Phi = H \cdot A$. Complex structures and fast dynamics in an active region result in the formation of structures that can be qualitatively described as twisted, sheared, *etc.* Such structures provide evidence for a strong departure from potentiality, the stored excess energy in the magnetic field and presence of major current systems in them (Chumak *et al.*, 2005).

Solar flares have been classified into two types, such as LDE flares *vs.* impulsive flares, or two-ribbon *vs.* simple loops. The former has often been thought to be explained by the so called "CSHKP" (Carmichael--Sturrock--Hirayama--Kopp--Pneuman) reconnection model, whereas the latter has been attributed to different models, such as the emerging-flux-reconnection model (Shibata, 1998). *Yohkoh*, however, has revealed that there are many common features in both types of flares, *e.g.,* the ejection of hot plasma (Shibata *et al.* 1995; Ohyama and Shibata, 1997); X-type or Y-type morphology suggesting the presence of current sheets or neutral points (Tsuneta *et al.* 1992; Tsuneta, 1997) and, change of field configuration, *etc.* In this view, it is now not easy to classify the flares into two types, and thus *Yohkoh* observations require a more unified view of flares. Shibata (1996, 1998) introduced a general model, known as the "unified model", to interpret all classes of flares. In his model, reconnection is caused by a plasmoid eruption and is observed in all classes of flares. These plasmoid eruptions may be mass ejections seen in the chromosphere such as a surge, spray, or filament eruption, or CMEs seen in the corona and beyond. However, in a study carried out by Subramanian and Dere (2001) on the source regions of 32 distinctly identified coronal mass ejections, they found that 41% of CMEs are not associated with any prominence eruption; rather they are associated with emergence or cancellation of magnetic fields in the active region. On the other hand 59% CMEs were found to be associated with a prominence eruption in the same or in a remote active region. This may reveal that CMEs may induce reconnection and the observed flares may be a result of such reconnection. However, many flares are not found to be associated with CMEs and thus reconnection through a plasmoid is an unlikely possibility. On the other hand, almost all flares are found to be associated with a small- or large-scale mass eruption, but not necessarily CMEs. Thus we may conclude that the primary energy for these manifestations is derived from the magnetic field of the active region but it is not known what kinds of magnetic-field variations are responsible for triggering such energetic phenomena. Therefore in this article we wish to address this question by studying the NOAA Active Region 8038 for the interval 07 to 13 May 1997 during which it produced a flare and associated CME on 12 May 1997. The observations, analysis, and results are presented in Section 2. We discuss results in Section 3 and offer conclusions in Section 4.

## 2. Observations, Analysis and Results

2.1 Magnetic field evolution:

Shown in Figure 1 is a sequence of a few high resolution magnetograms of NOAA Active Region 8038 made by the MDI instrument onboard the SOHO mission during 11--12 May, 1997. A detailed study of the evolution of magnetic field in the active region using a movie of total 60 re-registered magnetograms during 10--13 May, 1997 was conducted and presented earlier by Bharti *et al.* (2005). Therefore in this investigation we briefly describe salient features of the magnetic field evolution as follows.

i) Regular but discrete magnetic-field variations in the active region were taking place during the period under study in terms of emergence and cancellation of flux of both polarities, but predominantly of following polarity. Previously, while studying this active region using video-magnetograph observations from the Udaipur Solar Observatory (USO) and SOHO/MDI, Mathew and Ambastha (2000) showed appearances of emerging fluxes in this active region.

ii) Negative flux region S2 (*cf.* Figure1), the following polarity in the active region, was continually approaching the main positive leading polarity since 10 May, with resulting flux cancellations. However, the region S1 of following polarity (*cf.* Figure 1) was receding from the major leading polarity. In addition to the appearance of an emerging flux region (EFR) in S2, the magnetic fluxes were evolving, fragmenting, and merging indicating redistribution of fluxes in the active region (predominantly in trailing part). We confirmed appearances of EFRs in S2 as earlier reported by Mathew and Ambastha (2000) and Bharti *et al.* (2005). We found that among several EFRs observed in the active region, two EFRs of following polarity formed around 20:00 UT on 11 May, one each seen in S1 and S2. We believe these were of specific importance in the production of the solar flare on 12 May 1997.

iii) Moving magnetic features (MMFs) of north polarity were ejecting out from the major leading polarity (sunspot) as shown in Figure 1 with a quasi-periodicity of about ten hours. A total seven MMFs were observed during 10--13 May. However, out of these seven MMFs five were seen during the interval from 00:04:30 UT on 10 May to 20:52:30 UT on 11 May. The other two were seen at 22:28:30 on 12 May and 06:28:30 UT on 13 May. We found the plane-of-the-sky velocity of these MMFs in the range of 200--500 m s$^{-1}$.

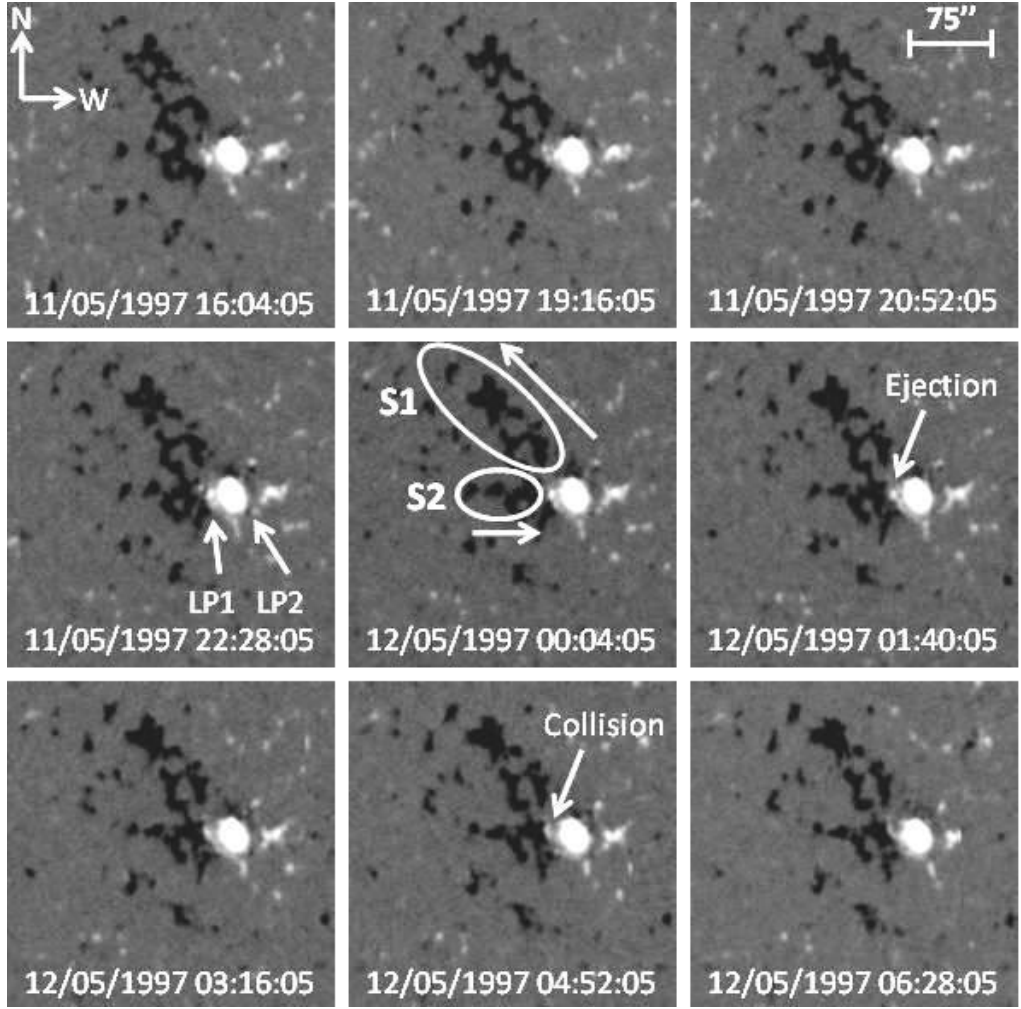

Figure 1: Sequence of a few selected high-resolution magnetograms of NOAA AR 8038 obtained by SOHO/MDI for the period 11--12 May 1997. The regions S1 and S2 are of following polarity, which are moving away and towards the leading polarity (sunspot), respectively. The ejection of north polarity flux from the sunspot is also shown. The onset of collision of north polarity flux, ejected from sunspot on 11 May around 20:52:30 UT, with a south polarity EFR in S2 began around 03:16 UT. The south polarity EFR in S1 is also intensified on 12 May.

iv) As shown in Figure 1, one such moving magnetic feature of north polarity flux (LP1), observed around 20:52:05 UT on 11 May began to collide with newly emerged south polarity flux in the S2 region that was approaching the major leading polarity (sunspot) around 03:16 UT on 12 May. In view of the poor temporal resolution of a few tens of minutes for magnetograms as well as CME observations (described in Section 2.3) it is not possible to arrive at a precise initiation time of the CME nor the collision time of the two opposite polarity fluxes observed in two magnetograms taken at 03:16 and 04:52 UT. However, we may estimate the initiation time of the CME within the temporal resolution limits of MDI and EIT as supported by signatures seen in EIT images. This collision of opposite polarity fluxes, *i.e.* cancellation of magnetic flux, is specifically important with regard to the production of the CME and the flare. About ten hours later to subsequent of collision in S2 the emerging flux regions were subsidizing. Conversely, the EFR of south polarity flux in S1 was significantly intensifying.

The above observations of the magnetic-field configurations of NOAA AR 8038 in terms of emerging flux regions, moving magnetic features, and cancellation of fields on short and long time scales provide a morphological picture of the ongoing process of magnetic-field variations in the active region. However, in order to understand among these factors of magnetic-field variations, which were predominant and responsible for triggering the flare and associated CME on 12 May 1997, we need a quantitative study of magnetic-field variations in relation to energy release by the active region. In next Section 2.2 we describe a quantitative estimation of magnetic-field variations.

2.2 Measurement of Magnetic-Field Variations:

2.2.1 Measurement of Magnetic Flux:

The magnetic flux [$\Phi$] is estimated as the product of magnetic-field strength [$H$] and area [$A$] of a given active region as presented in the following relation (1).

$$\Phi = H.A \qquad (1)$$

$H$ is estimated using SOHO/MDI 96-minute full-disk observations with the help of functions provided in **SolarSoft**. The level-1.8 data from http://soi.stanford.edu/magnetic/index5.html are processed for further measurements. We employ the Carrington Projection method for foreshortening correction using the `map_carrington.pro` routine provided in **SolarSoft**. The magnetic flux is estimated separately for regions of leading and following polarities identifying them distinctly as discussed below. The regime (pixels) having magnetic field strength $\geq$ 10% of the positive peak magnetic-field strength is considered as a leading polarity region. Similarly, the regime (pixels) belonging to the magnetic-field $\leq$ 10% of the negative peak magnetic field strength is considered as a following polarity region. The rest of the region is considered as background or quiet Sun. The size of the pixel is 1.97 arcsec, and therefore the area occupied by both the polarities is then estimated separately by total number of pixels multiplied by area of single pixel, *i.e.* $725^2 \times 1.97^2 \times 10^{10}$ cm$^2$. The magnetic flux [$\Phi$] for the respective polarity is then estimated by multiplying the corresponding total magnetic-field strength with the above estimated area (*c.f.* Equation 1), and is plotted for each day as shown in Figure 2 (top panel).

2.2.2 Measurement of Magnetic-Field Gradient:

The separation between leading and following polarities is considered as the distance between their respective centers of mass. The center of mass coordinates ($x_c$, $y_c$) of both the polarities are obtained with the aid of the following equations

$$x_c = \frac{\sum_{i,j} x(i,j) B_{los}(i,j) ds}{\sum_{i,j} B_{los}(i,j) ds} \qquad (2)$$

and

$$y_c = \frac{\sum_{i,j} y(i,j) B_{los}(i,j) ds}{\sum_{i,j} B_{los}(i,j) ds} \qquad (3)$$

where $B_{los}(i,j)$ is the line of sight magnetic field strength corresponding to the pixel location having coordinate [$x(i,j)$, $y(i,j)$] and d$s$ = dxdy is the area of each pixel. The summation runs over the total number of pixels in the region of interest. The estimated center of masses of leading and following polarity is then employed in the following equation to obtain the polarity separation (d$z$).

$$dz = \sqrt{(x_{cl} - x_{cf})^2 + (y_{cl} - y_{cf})^2} \qquad (4)$$

Where, ($x_{cl}$, $y_{cl}$) are the coordinates of the center of mass of the leading polarity and ($x_{cf}$, $y_{cf}$) are of the following polarity. The magnetic field gradient is then estimated by dividing the difference of total magnetic-field strengths of leading and following polarities [dH] with the above estimated polarity separation [d$z$]. The variation of the estimated magnetic field gradient (d$H$/d$z$) over the passage of the active region on the disk is shown in Figure 2 (middle panel).

2.2.3 Measurement of angular rotation of leading polarity:

The leading polarity LP2 is elliptical in shape (*c.f.* Figure 1). Its orientation can be described as an angle between its major axis and the Equator in the anti-clockwise direction. This angle is defined as the rotation angle of the sunspot (Zhang *et al.*, 2008). In order to estimate the rotation of the leading polarity, we employ the fit_ellipse.pro procedure written in IDL and provided by the Coyote IDL programming package. The structure corresponding to the leading polarity is identified as the region (pixels) having magnetic field strengths (H) ≥ 10% of the peak magnetic field strength. This condition is the same as the condition defined for identifying the structure of the polarities as described in section 2.2.1. The fit_ellipse.pro procedure is then applied on the identified region in order to obtain the parameters of the best fitted ellipse on the region. The output parameters of this procedure include the major and minor axes as well as their orientation from the Equator of the Sun in an anti-clockwise sense. The orientation angle of the major axis therefore gives the information of the rotation of leading polarity as shown in Figure 2 (bottom panel).

The eight hours averaged parameters *viz.* magnetic flux, gradient, and rotation are shown in Figure 2. The error in each parameter is estimated as standard deviation of five 96-minute observations comprising one eight hour data set, and is shown in each panel of Figure 2. Top panel of the Figure shows the time variations of eight hour averaged magnetic fluxes of leading and following polarity. The variations show that the magnetic fluxes in the leading and following polarities have started to build from 7 May 1997. They attained a maximum value on 11 and 12 May 1997, however, it started to decrease afterwards. The middle panel of Figure 2 shows the time variation of the eight hour averaged magnetic field gradients. The variation shows that magnetic-field gradient was significantly higher around 07 May but then it decreased until 10 May. The gradient started to increase on 11 May and attained a maximum value on 12 May 1997 -- the day of CME/flare event. After the flare, the gradient started to decrease and continued to lower levels. The bottom panel of Figure 2 shows the temporal variation of eight hour averaged rotation angle of leading polarity. The variation indicates that the leading polarity was rotating anti-clockwise with $0.25°$ $h^{-1}$ from 7 May 1997 until 01:40:05 UT on 12 May when the rotation angle reached to $\approx 100°$. This finding is in contrast to Li *et al*. (2010) who reported minor clockwise rotation of $1°$ $h^{-1}$ during 09--12 May 1997. We further found that the magnetograms beginning at 03:16:05 UT on 12 May showed a reversal of rotation in the leading polarity.

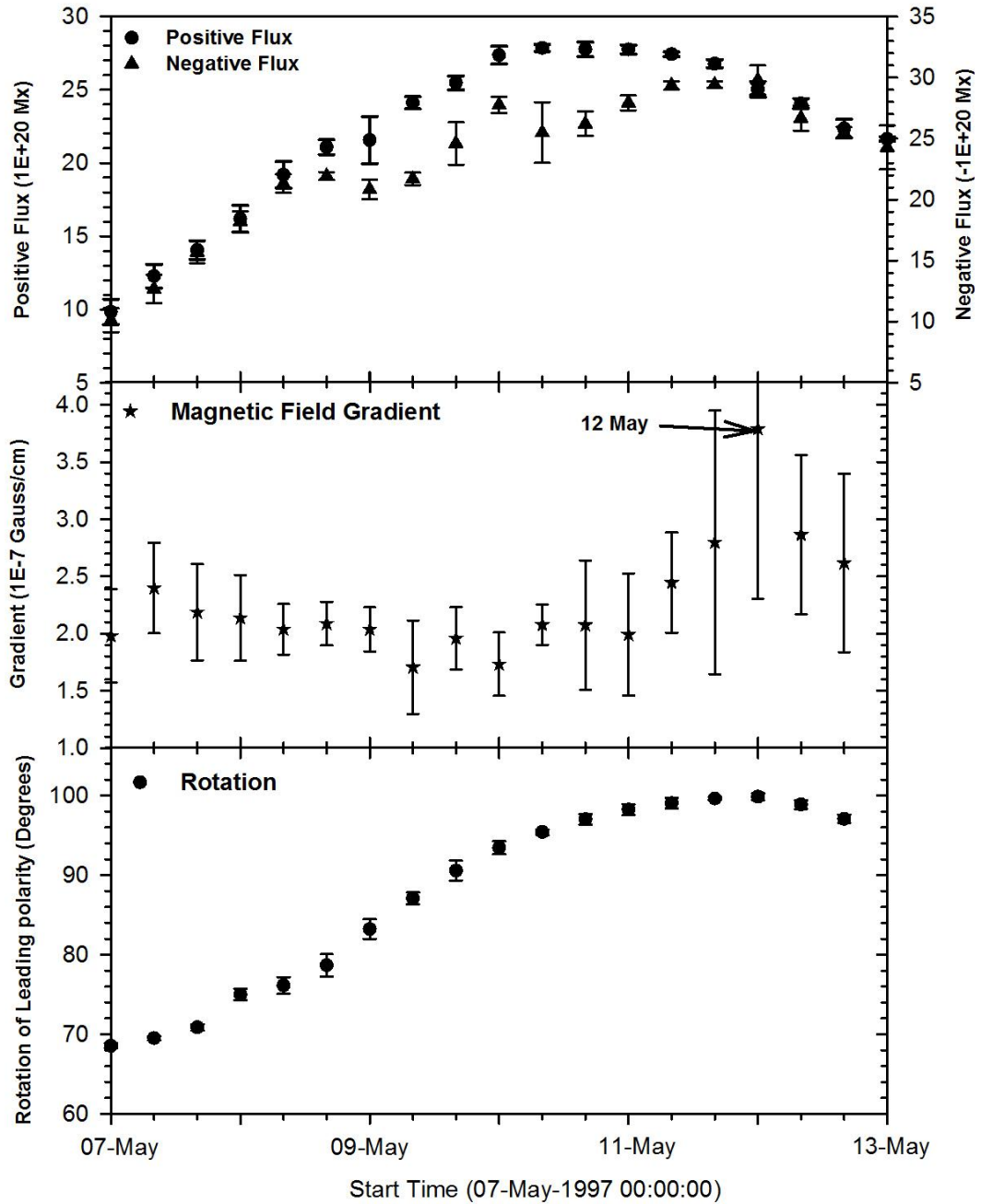

Figure 2: Temporal evolution of eight hour averaged magnetic fluxes of leading and following polarities of NOAA 8038 (top panel), the gradient (middle panel), and the rotation angle of leading polarity (bottom panel).

2.3 The Hα Observations:

2.3.1 The Chromospheric Activity During 10--13 May 1997:

Shown in Figure 3 is a sequence of high resolution Hα filtergrams obtained from USO during 10--13 May 1997. It may be noted that the filament that stretched over the active region and was visible on 10 May, was disrupted and was not visible on 11 May at 03:47 UT. However, it reformed around 06:30 UT on 11 May. This filament showed considerable activity in the active region during this period. In addition to filament activity, the plage intensity was also varying in the active region particularly during the period of appearance of emerging flux regions (EFR). The filament again appeared disrupted on 12 May from the beginning of our observations at 03:03 UT, which appears to be the potential candidate to brightening the EIT loops (*cf.* Section 2.4 and Figure 7 – right panel). However, as shown in Figure 3, we could see several dark and bright surges from 03:03 to 04:40 UT on 12 May. It may also be noted that plage areas, from following polarity fluxes, East of the dark leading sunspot, brightened up on 12 May. Such chromospheric activities observed beginning on 10 May in the active region is consistent with the appearances of EFRs and their growths and decays as seen in MDI observations.

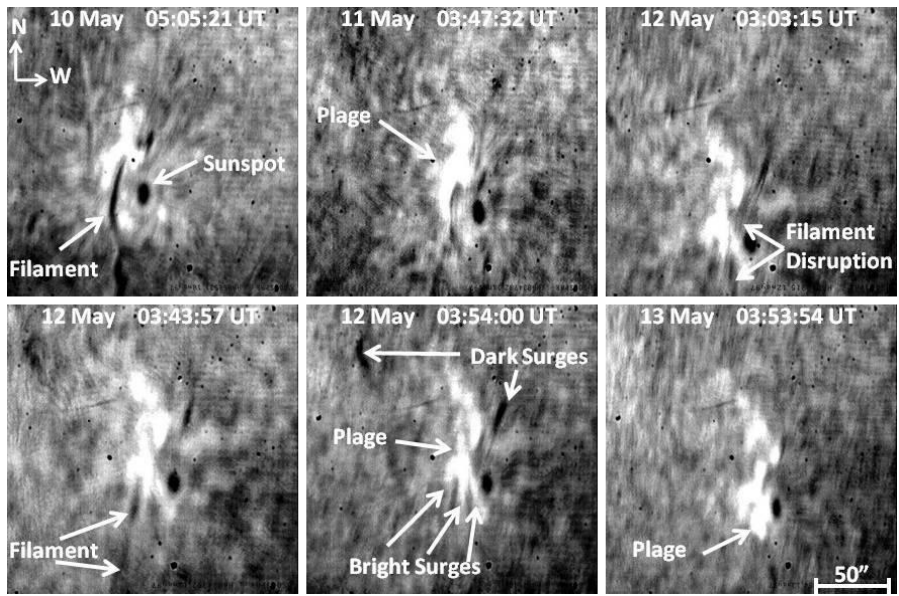

Figure 3: High resolution Hα filtergrams, one for each day for the period 10--13 May, 1997, showing bright and dark mass ejections and considerable filament activity and plage intensity variations in the active region. Significant filament activity may be noted between 03:03 and 03:54 UT.

2.3.2 The Filament Activity on 12 May 1997:

Solar flares are associated with magnetic neutral lines where the vertical component of photospheric magnetic-field changes sign. These neutral lines are often marked by filaments, which become turbulent and disappear before the flare occurs (Van Tend and Kuperus, 1978). We investigate the role of the filament with the flare and CME that occurred on 12 May 1997 employing the Hα observations taken at a cadence of 30 seconds using the US Air Force Solar Observing Optical Network (SOON). Figure 4 shows the temporal sequence of a few selected Hα images. The filament appearing as a dark curvilinear line and extending from North to South of the active region has been designated in four parts *viz.* F1, F2, F3, F4 as shown in Figure 4 (04:40:09 UT). This categorization is made based on their activity seen during the evolution of the region particularly before the occurrence of the flare and CME on 12 May 1997. For example, the part F1 is not visible at 04:29:09 UT, but it appeared at 04:34:09 UT and again disappeared at 04:35:09 UT. We observed frequent disruptions of segment F1 and material moving out in the plane of the sky towards the Northwest above the leading sunspot LP2. This activity appears to be associated with frequent appearances of emerging fluxes of following polarity in the S2 region and flux cancellations with positive polarity. However, the disruption of the filament leads to upward motion of the filament due to enhanced gradients at the site of the filament. The magnetic-field gradient was observed to increase beginning on 11 May 1997 (*cf.* Figure 2). On the other hand, the longer F2 part of the filament, which was stable until 04:41:09 UT, started to detach from F3 around 04:41:39 UT. The breakup and eruption of F2 appears to be associated with the flare and CME. This time is also confirmed by GOES X-ray flux observation as the start time of flare (*cf.* Figure 5b). The F3 and F4 parts of the filament, which almost remain connected with F2 in the pre-flare stage, significantly erupted at 04:43:39 UT, around the Hα flare time described in next Section 2.3.3

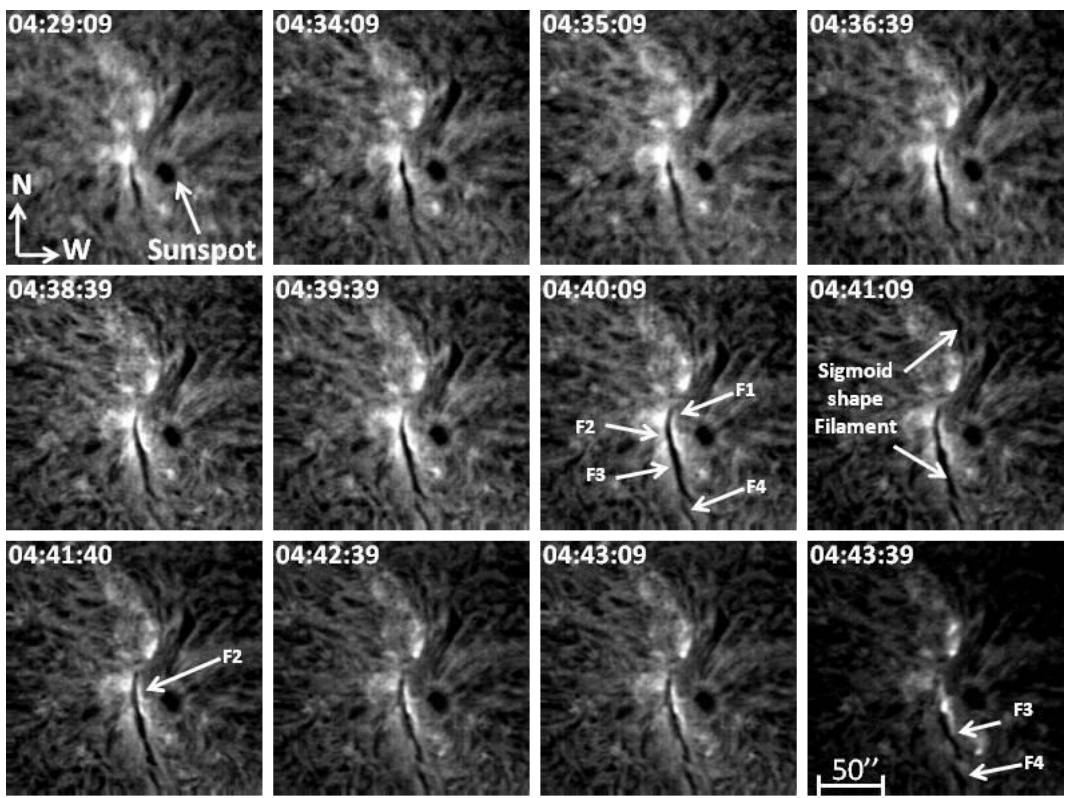

Figure 4: Sequence of Hα filtergrams showing filament activity on 12 May 1997 prior to CME--flare eruption. The F1 part of the filament shows frequent activity, while F2 and other parts were disrupted before the flare and CME around 04:43 UT. The filament also shows reverse-shaped sigmoidal structure.

### 2.3.3 The Hα Flare:

In Figure 5a we show a sequence of Hα flare evolution observed by SOON (Section 2.3.2). The Hα flare was first observed at 04:43:39 UT. The temporal cadence of observations was 30 seconds, and therefore the onset of the flare was between 04:43:09 and 04:43:39 UT. However, the GOES flare onset time in 1--8 Å was at 04:41:21 UT, consistent to F2 disruption. Nevertheless, taking 5σ enhancement above background as shown in Figure 5b, we may confidently consider X-ray flare onset at 04:42:36 UT in 1--8 Å. On the other hand, the hard X-ray flare (25--50 keV) observed by BATSE, shown in Figure 5c, started significantly later at 04:50 UT. It might be possible that the instrument sensitivity limited observations in the early phase. However, this event is found associated with interplanetary protons >10 MeV. We plan to present results of hard X-ray, MW, and protons in another article. At both ends of the flare strands the bright material shot out as seen in the frame taken at 04:45:09 UT (Figure 5a). If these bright blobs are joined together then indicating that the mass ejection was of a spherical waveform resembling the blast wave seen by SOHO/ EIT at the initial phase of the CME. The speed of these bright blobs was in the range 200 to 300 km s$^{-1}$. The flare was composed of several bright/eruptive centers forming the ribbon structure. The ribbon structure in the flare was very clear beginning at 04:44:39 UT. The separation between the flare kernels A and B of the two opposite ribbons, shown in the frame taken at 04:50:39, has been measured as a function of time and is presented in Figure 6. The observations revealed that in the initial phase (04:44 -- 04:46 UT) these kernels were separating away from each other with the speed of >80 km s$^{-1}$, which, however, slowed down exponentially to ≈ 20 km s$^{-1}$ after 04:48 UT. Further, we estimated the reconnection electric-field (**E$_R$**) variation during flare as follows. We estimated the magnetic-field strength with the time cadence of 30 seconds with the aid of cubic spline interpolation technique employing the 96-minute full-disc MDI magnetic-field observations. The cubic spline interpolation algorithm for finding the interpolated magnetic field ($B_{int}$) is based on the smoothed interpolation described by Press et al. (1992) in their book *"Numerical Recipes in C: The Art of Scientific Computing"* considering first and second order derivatives as following.

$$B_{\text{int}}(t) = A \cdot B(t_j) + C \cdot B(t_{j+1}) + D \cdot B''(t_j) + E \cdot B''(t_{j+1}) \qquad (5)$$

Here B($t_j$) and B($t_{j+1}$) are the observed values of photospheric magnetic field at time $t_j$ and $t_{j+1}$ respectively and $t$ is the intermediate time for which magnetic field is to be estimated. A, C, D, and E are the parameters defined as following:

$$A \equiv \frac{t_{j+1}-t}{t_{j+1}-t_j}; \quad C \equiv 1 - A = \frac{t-t_j}{t_{j+1}-t_j};$$

$$D \equiv \frac{1}{6}(A^3 - A)(t_{j+1} - t_j)^2 ; \text{ and } E \equiv \frac{1}{6}(C^3 - C)(t_{j+1} - t_j)^2$$

Te estimated $B_{int}$ is then multiplied with the corresponding ribbon-separation speed (V) so as to obtain the reconnection electric field ($E_R$) from the following relation.

$$E_R = V \bullet B_{int} \quad (6)$$

The ribbon separation speed as well as the reconnection electric field mimics the passage of the CME. The flare kernel B of the west ribbon struck the leading sunspot around 04:54:39 UT and stopped further movement.

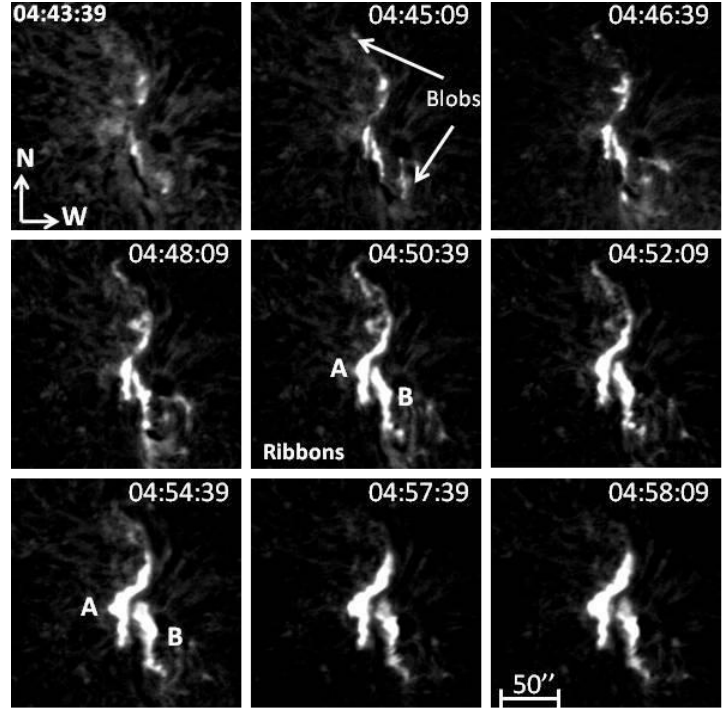

Figure 5a: A sequence of high-resolution Hα filtergrams of the 1B flare observed in NOAA 8038 on 12 May 1997 by SOON. The bright mass ejection at both ends of flare ribbons may be seen at 04:45:09 UT. The flare ribbons move away from each other which may be noted by the separation of flare kernels A and B.

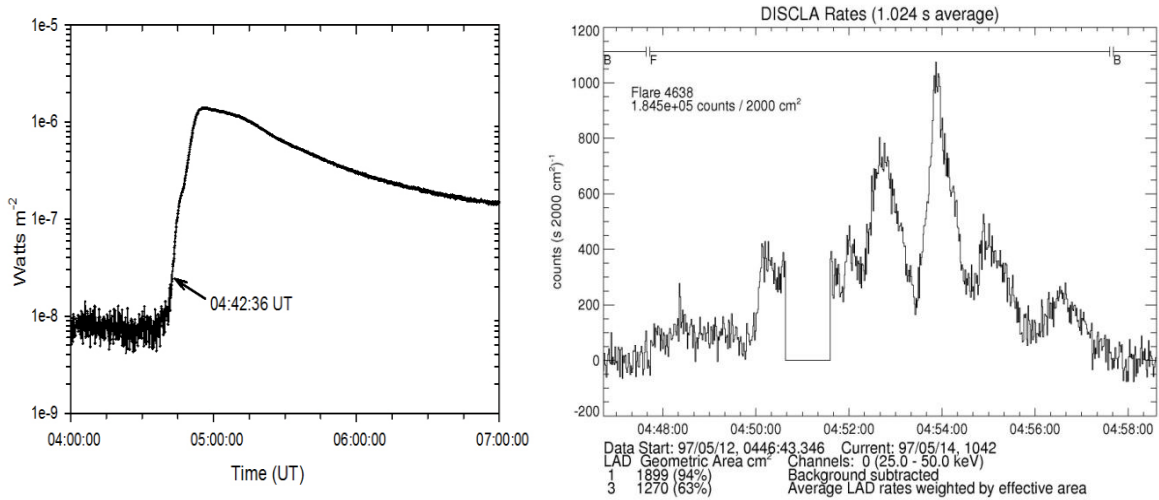

Figure 5(b, c): The three-seconds X-ray flux measurements in 1--8Å by GOES 12 shows onset of the 12 May 1997 flare at 04:42:36 UT taken to be the 5σ enhancement above pre-flare background, while the hard X-ray in 25--50 keV (right panel) shows the onset around 04:50 UT.

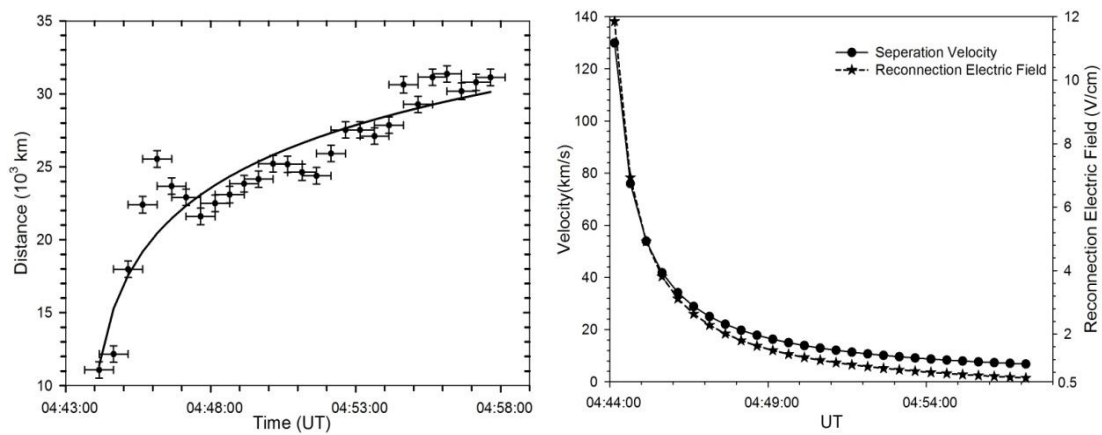

Figure 6: Left - The separation of flare kernels A and B of two opposite flare ribbons (*cf.* Figure 5a) over time. The velocity decays exponentially from 120 to 10 km s$^{-1}$ in a ten minute interval. Right – The deceleration of flare ribbons and the reconnection electric-field variation over time.

2.4 The Coronal Mass Ejection (CME):

According to Thompson *et al.* (1998) and Plunkett *et al.* (1998), beginning at about 04:35 UT, EIT recorded several CME signatures, including dimming regions close to the eruption, post-eruption arcade formation, and a bright wave-front propagating quasi-radially from the source region. However, the CME in the form of an EIT wave as shown in Figure 7 (left) was first observed in the difference images in 195 Å between 04:34 and 04:50 UT suggesting that the disturbance began at a time between these two times with an uncertainty of less than 18 minute. The CME was later observed by the *Large Angle Spectrometric Coronagraph* (LASCO) as a "halo" CME. In this investigation we consider analysis of plane-of-the-sky images of magnetograms, the Hα flare and the EIT images. We do not attempt to transfer or extrapolate a full 3D analysis. The CME emerged from its heliographic location N23 W07 and traveled outward as a spherical wave (*cf.* Figure 7 - left) with a speed of about 250±20 km s$^{-1}$ (Plunkett *et al.*, 1998). In Figure 7 (right) we further show the EIT 195Å plane-of-the-sky images. We note explicitly that around 03:08 UT the first EUV emission started to brighten in the core of the reversed-S sigmoid configuration, which implies that slow magnetic reconnection was taking place there (Cheng *et al.*, 2010) perhaps initiated as a consequence of the filament disruption observed on 12 May at 03:03 UT (*cf.* Figure 3). We further note that as the field lines in the reversed-S sigmoid configuration continually reconnected, the filament (flux-rope) rose slowly and disrupted which unambiguously reveals the first brightening of the coronal loops around 03:25:11 UT (shown by arrow). The brightening of the coronal loops continued. As seen at 04:34:50 UT, all loops brightened, in close temporal agreement with the disruption of part F1 of the filament where mostly flux cancellation was taking place. In comparing this EIT image with the difference image (04:34 - 04:16 UT) in the left panel at 04:34 UT, we find a signature of dimming in the region but not the signature of blast wave. We consider this time as the onset of the filament rising due to loss of equilibrium in the active region (*cf.* Section 2.3.2). We believe the rising filament caused the dimming and then it continued to move up into the corona. Filament eruptions from the Sun are often accompanied by a dimming of the local coronal emission at many different wavelengths and appear to be associated with transient coronal holes (Harrison *et al.*, 2003; Harra *et al.*, 2007; Imada *et al.,* 2007; Reinard and Biesecker, 2008, 2009; Jin *et al.*, 2009; Dai *et al.*, 2010, Robbrecht and Wang, 2010). The dimmings are in most cases caused by a decrease in the coronal density due to the opening up of the magnetic field and escape of the entrapped material into the heliosphere. The closing down of the flux proceeds from the inside outward, with the field lines rooted nearest to the photospheric polarity inversion line (PIL) pinching off first, giving rise to a progressively growing post-eruption loop arcade (Kopp and Pneuman 1976). In the present case it appears that the disruption of the F2 part of the filament due to continuous loss of equilibrium (because of increasing gradients and rotation angle of LP2) initiated the flare and the CME starting around 04:43 UT as seen in GOES 1-8 Å and Hα observations. The CME can be seen in the difference image at 04:50 UT due to 18 min cadence of the EIT instrument. Thus, in contrast to Thompson et al., (1998) and Plunkett et al., (1998) we do not consider the onset of CME at 04:35 UT, rather we propose that the CME and flare began together around 04:42:36 UT (onset of the X-ray flare) as a consequence of opening of high altitude loops and consequently their reconnection. The filament detached from the photosphere and started to rise to the corona through the chromosphere perhaps with projected speed of 250±20 km/s measured by Thompson et al., (1998) from EIT difference images, and by us from the bright mass ejection observed in Hα. The rising filament motion caused continuous brightening of EIT loops as it was moving up into the corona starting from 03:25 UT seen in Figure 7 (right) and finally opened the loops around 04:42:36 UT so as to generate EIT wave seen later at 04:50 UT in EIT difference images. Due to the slow cadence of EIT image acquisition the reconnection time and hence the wave onset times were missed.

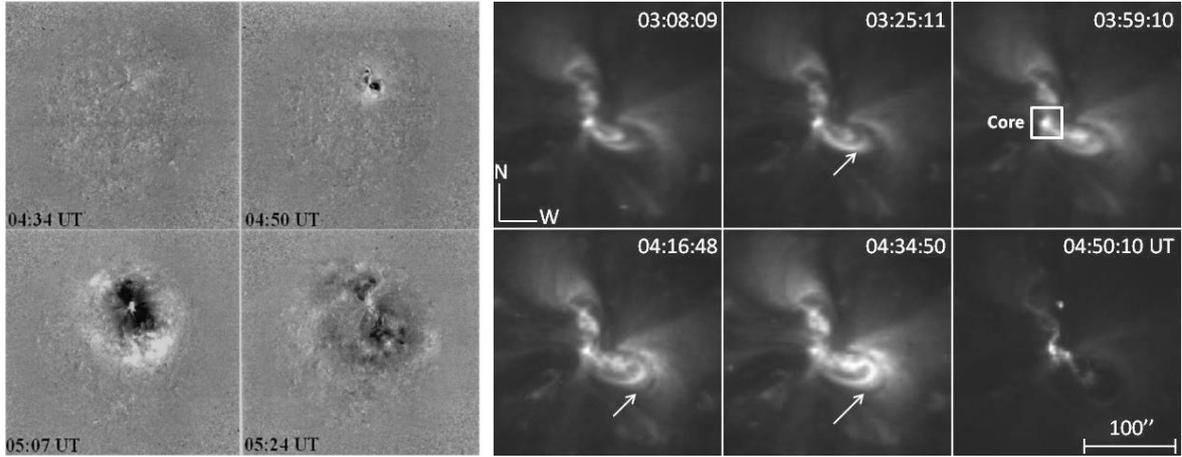

Figure 7: Left - SOHO/EIT 195Å running-difference images of large-scale wave transient moving out as a CME. The wave-like structure began around 04:34 UT. Right – The time-series of SOHO/EIT 195 Å images. Loop brightening may be noted at 03:25:11 UT, which continued to brighten the whole active region at 04:34:50 UT.

## 3. Discussions

We carried out a detailed study of the evolution of NOAA AR 8038 during its passage on the disk from 05 to 16 May 1997. In Figure 8, we attempt to use a schematic model to explain the full evolution of the flare--CME from early development to the ultimate eruption. It is well established that the full evolution of the CME is divided into four phases: i) the build-up phase, ii) the initiation phase, iii) the main acceleration phase, and iv) the propagation phase (Cheng, Ding, and Zhang, 2010). However, our current investigation is focused on the first three components of the CME, which include the flare and CME as two modes of a single energy release system.

The build-up phase lasted five days, from 07 to 12 May 1997, before the CME. As discussed earlier, it is characterized by many pre-cursor signatures: flux cancellation, filament activity, sigmoid, and Hα plage and EUV brightening, even though these signatures are neither necessary nor sufficient for an eruption. Formation and evolution of filaments have been extensively studied for many years. However, our observations of the filament formation and activation using two flux systems driven by the convergence of opposite polarities along the PIL confirms the simulation study carried out by Welsch *et al.* (2005). Chae *et al.* (2001) and Bharti *et al.* (2007) proposed that slow magnetic reconnection driven by converging motions may occur at all times in the chromosphere. The continuous reconnection can result in both the overlying field lines straddling the neutral line and the low-lying, core field lines (Chae *et al.*, 2001; Welsch *et al.*, 2005).

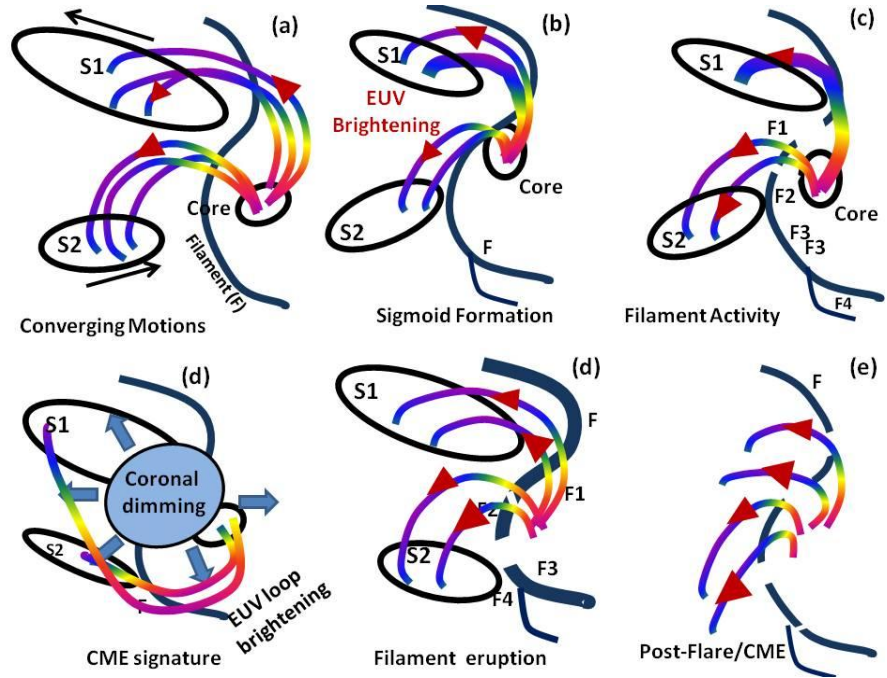

Figure 8: A schematic drawing of the evolution of the active region in the framework of Flare--CME event. Rainbow colored lines refer to the overlying magnetic-field lines. The leading and the following polarity regions are demarcated by black ellipse-shaped structures shown by core and S1 as well as S2 respectively. The dark blue line represents the filament overlying the polarity inversion line (PIL). The direction of magnetic-field lines is shown by magenta arrows. The light-blue arrows (d) represent the propagation of coronal dimming.

Furthermore, some EUV brightening/jets and small Hα eruptions/surges took place at the site of magnetic-flux cancellation (Liu, 2004). We propose that as the positive and negative fluxes moved close to each other near the PIL the anti-parallel inner ends of the two bundles of the loops reconnected slowly and continuously in the lower atmosphere (*i.e.* the chromosphere). These convergence motions almost perpendicular to the PIL (Cheng, Ding, and Zhang, 2010; Li et al., 2010) started to form sigmoidal structure (Figure 8a). At the same time, it also produced the low-lying field lines that are nearly parallel with the PIL. As time progresses, the lower field lines in the eastern part and the pre-existing field lines in the western part, being both J-shaped, moved closer to each other, driven by the continuous convergence motion along the PIL and formed a reversed-S sigmoid structure in the projection sky plane (Figure 8b). We conjecture that the ends of the two bundles of the J-shaped loops, on the opposite sides of the PIL, reconnected as tether cutting and formed the little twisted field lines, while the energy released through the reconnection heated the plasma in the middle part of the reversed-S sigmoid configuration, thus producing the EUV core brightening at 03:08 UT. The shortest field lines submerged into the sub-photosphere after the slow reconnection, which was manifested by the magnetic cancellation in the photosphere, and possibly produced the filament activity as seen in the chromosphere (Figures 8b--c). Recently, Tripathi *et al.* (2009) and Green and Kliem (2009) also reported such a sigmoid structure coming into existence after a pair of J-shaped arcs reconnected through flux cancellation in the photosphere. We propose that these sigmoid structures provide observational evidence of the flux rope existence prior to the flare-CME eruption. The active filaments condensed at the dip of these two J-shaped field lines, as indicated in Figure 8c. As the filament mass flew down from 10 May onwards along the field lines, more field lines rose and served as the overlying loops. These loops were heated slowly and remained invisible in the EUV until about an hour prior to the CME eruption on 12 May 1997. It is further proposed that as long as some open field lines exist at the reconnection site, part of the filament mass may erupt as EUV jets. However, although part of the filament mass (F1) flew down slowly or erupted, the rest of the filament appeared to be quite stable in the dip of the field lines at all times prior to the CME eruption (*cf.* Figure 4) as shown in Figure 8(c--d).

The initiation phase occurred when the upward force within the sigmoid was able to overcome the tension force of the overlying field lines. As more and more J-shaped loops reconnected by tether cutting, the twisted field lines in the reversed-S sigmoid configuration, beneath the overlying loops, moved up due to an increased upward magnetic hoop force and a decreased downward magnetic stress (Moore *et al.*, 2001; Liu *et al.*, 2007; Sterling *et al.*, 2007). The rising, twisted field lines pushed the overlying loops upward. When the overlying loops were stretched to a certain extent due to the tether-cutting reconnection, a current sheet between the legs of the distended overlying field line was formed under the loops so that a fast, runaway reconnection was subsequently initiated, leading to the main energy-release phase and the impulsive acceleration of the CME; this is the standard model of eruptive flares (Hirayama, 1974). Another possibility leading to the main eruption is the triggering of MHD instability of the flux rope formed from the tether-cutting reconnection, through the kink and/or torus instability (Török and Kliem 2005; Kliem and Török 2006, Démoulin and Aulanier, 2010).

The subsequent main acceleration phase is believed to be caused by runaway magnetic reconnection, coupled with the explosive poloidal flux injection into the rising flux rope.

Our observations show a fast reconnection rate (high speed of ribbon separation) and high reconnection electric fields in the main phase suggesting that the reconnection rapidly injects a large amount of poloidal flux into the twisted field lines, thus supplying a stronger upward driving force so as to impulsively accelerate the CME flux rope. On the other hand, the CME eruption led to a decrease in the magnetic pressure below the flux rope, which caused a faster inflow toward the current sheet and enhanced the runaway reconnection. This positive feedback process effectively released the magnetic free energy stored in the lower corona, which was converted into the kinetic energy of the CME and also produced the enhanced soft X-ray emissions (Li *et al.*, 1993). On the contrary the HXR emission in 25--50 keV does not appear to be directly associated with the main phase of the flare as it is delayed by almost seven minutes from the onset of soft X-ray and Hα flare. It is possible that the instrumental sensitivity of BATSE might have restricted the early detection of HXR emission. Our proposal is strengthened by observed protons of >10 MeV in association with this event, which, must have been accelerated along with the CME in the inner corona. However, this event failed to reach the NOAA Space Weather Prediction Centers' (SWPC) criteria to assure its place as a "qualified" proton event. On the other hand, the fact that interplanetary protons were seen at Earth is important. The hard X-ray, microwave, and interplanetary particle aspects of this event are beyond the scope of this paper but will be presented in a future publication. The other possibility might be that HXR emission is more associated with the CME propagation in the outer corona. This suggests early acceleration of the CME (in the inner corona) accelerated electrons to high energy during its propagation outwards and those runaway high energy electrons produced the hard X-ray emission. In the present case, a significant portion of the energy was carried by CME so the input energy for the flare was keep on decreasing over time as exhibited by reconnection electric field (see also Zhang *et al.*, 2004; Qiu *et al.*, 2004; Temmer *et al.*, 2008). It suggests that the main acceleration phase of the CME in the inner corona is likely to be caused by fast runaway magnetic reconnection. Later on, the CME propagated with an almost constant velocity in the outer corona. In general, these observational results are consistent with the standard CME-flare model.

Moreover, EUV loops brightened around 04:34 UT in association with filament eruption activity. This led to a depletion of mass in the lower atmosphere near the active region and formed a coronal dimming (Thompson *et al.*, 1998, Harra and Sterling, 2001) a few minutes before the main CME-flare eruption at 04:43 UT. As the magnetic reconnection progressed, the reconnection site rose gradually. The upward moving reconnection site induced the flare ribbons to separate horizontally at the base of the corona, as evident in Hα. Beneath the reconnection site, the newly reconnected magnetic loops were filled by the plasma that evaporated from the chromosphere and the sigmoid magnetic structure evolved into post-flare loop arcades (see also Liu *et al.*, 2007), as shown in Figure 8(f). After the main phase which lasted about seven minutes, the runaway reconnection came to a stop. The CME now entered into its simple propagation phase: the CME was propagating with a nearly constant speed or with a deceleration in the outer corona.

## 4. Conclusions

We studied the evolution of magnetic configuration for a few days prior to the flare and CME activity on 12 May 1997. During the build-up phase, we observed many precursors such as magnetic field cancellations, filament activation, bright and dark surges and plage brightening in the chromosphere, instantaneous EUV jets, and a reversed-S sigmoid structure. All the features were physically related to a persistent, slow, magnetic reconnection in the lower solar atmosphere, which was manifested as photospheric magnetic cancellations. Before the flare and CME eruption on 12 May 1997 there was a long period of reconnection occurring in the lower layers, resulting in the transfer and accumulation of magnetic free energy manifested as magnetic flux, as well as the formation of a magnetic structures favorable for eruption, *i.e.*, the sigmoid structure of this event derived from the rotation of the leading sunspot. In addition to the process of flux cancellation, the emergence of magnetic flux may also play an important role in transferring magnetic free energy from the sub-photosphere into the corona (Tian *et al.*, 2008; Archontis *et al.*, 2009). We observed converging motions of opposite polarities in this active region beginning on 10 May 1997. The magnetic-field shear can be caused by the converging motion of opposite magnetic fluxes (Titov *et al.*, 2008), which, in turn, produces the sigmoid-shaped structure as seen in Hα and EIT observations of 12 May 1997. The enhanced shear aided the accumulation of free energy in the corona. Therefore, the build-up phase accumulated sufficient magnetic free energy, beginning on 07 May 1997, for the eventual initiation and the final eruption of flare and CME event on 12 May 1997.

In the present event, the EUV emission started to brighten in the core of the reversed-S sigmoid configuration, which implies that slow magnetic reconnection was taking place there. As the field lines in the reversed-S sigmoid configuration continually reconnected, the F1 part of the filament (flux-rope) rose slowly and then erupted. Furthermore, the convergence of opposite polarities resulted in increasing magnetic-field gradients. We believe that the loss of equilibrium began in this configuration suggesting an onset of the eruption at 04:34 UT seen as emission of loops in EIT images. However, the inner-core magnetic reconnection prior to the eruption, combined with the bipolar magnetic structure in the active region and the absence of remote brightenings, seem to rule out the breakout model as the triggering mechanism of this flare associated CME event. Instead, we think that this eruption is most consistent with the tether-cutting initiation model.

**Acknowledgments** The authors sincerely acknowledge the Air Force Weather Agency (AFWA), who manage and operate the *Solar Observing Optical Network* (SOON) around the world, and for their support of the digital archive system known as SOONSPOT that provided the SOON data. We also acknowledge David Fanning for providing Coyote's Guide to IDL Programming. The Authors are thankful to anonymous reviewer for their useful suggestions.